\documentclass[5p,twocolumn,twoside,times,fleqn,number]{elsarticle}

\usepackage{amsmath}
\usepackage{amssymb}
\usepackage{amsfonts}

\newcommand{\bl}{\boldsymbol{\lambda}} 
\newcommand{\hypergeom}{F} 
\newcommand{\dop}{\textrm{d}}  
\newcommand{\timeconstant}{\nu} 
\newcommand{\Laplacetransform}{\mathcal{L}} 
\newcommand{\diagramset}[1]{#1} 
\newcommand{\singlestepset}[1]{\mathbb{#1}} 
\newcommand{\powerset}{P} 
\newcommand{\setJ}{J}

\journal{Chemical Physics}

\begin{document}

\begin{frontmatter}

\title{$1/f^\beta$ noise in a model for weak ergodicity breaking}

\cortext[cor]{Corresponding author. Tel.: +49 441 7983619; fax.: +49 441 7983080;
              current address: Carl von Ossietzky Universit\"{a}t Oldenburg, Institut f\"{u}r Physik, 26111 Oldenburg, Germany}

\author{Markus Niemann\corref{cor}}
\ead{markus.niemann@uni-oldenburg.de}

\author{Ivan G.~Szendro\corref{}}
\ead{szendro@pks.mpg.de}

\author{Holger Kantz}
\ead{kantz@pks.mpg.de}

\address{Max-Planck-Institut f{\"{u}}r Physik komplexer Systeme, Noethnitzer Stra{\ss}e~38, 01187 Dresden, Germany}

\begin{abstract}
In systems with weak ergodicity breaking, the equivalence of time averages
and ensemble averages is known to be broken. We study here the computation of
the power spectrum from realizations of a specific
process exhibiting $1/f^\beta$
noise, the Rebenshtok--Barkai model.
We show that even the binned power spectrum does not converge in the
limit of infinite time, but that instead the resulting value is a random
variable stemming from a distribution with finite variance. However, due
to the strong correlations in neighboring frequency bins of the spectrum,
the exponent $\beta$ can be safely estimated by time averages of this type.
Analytical calculations are illustrated by numerical simulations.
\end{abstract}

\begin{keyword}
$1/f$ noise \sep weak ergodicity breaking \sep spectral estimation

\PACS 05.10.Gg \sep 05.40.-a \sep 05.45.Tp
\end{keyword}


\end{frontmatter}

\section{Introduction}
In recent years, the study of long range correlations in experimental data has
moved into the focus of interest. Whereas a large amount of empirical results
has been obtained using detrended fluctuation analysis \cite{PengHavlinStanleyEtAl95}, 
the same information is, in principle, also contained in the power spectrum. 
Observations of nontrivial power law decays of the power spectrum have a long
tradition and are usually denoted by $1/f$-noise, or, more precisely,
$1/f^\beta$ with $\beta\approx 1$. Examples include noise in electric
resistors, in semiconductor devices (flicker noise), 
and fluctuations in geophysical data 
\cite{Weissman88,HoogeKleinpenningVandamme81,BalascoLapennaTelesca02}. 
 
By weak ergodicity breaking one describes the behavior of a system 
which is ergodic in the sense that a single trajectory is able to explore
completely some invariant component of the phase-space (it is indecomposable),
but where the invariant measure is not normalizable, so that the time scales
for this exploration might diverge (in other words, there is no finite
microscopic time scale). This concept goes back to Bouchaud \cite{Bouchaud92}.
A consequence of this is that 
time averages computed on a single (infinite) trajectory will not converge to 
the ensemble mean. Instead, the time average is a random variable itself, as a
function of the initial condition of the trajectory, which is drawn from a
distribution with non-zero variance. It has been shown that 
Continuous-Time Random Walks \cite{BelBarkai05}, but also specific deterministic dynamical
systems \cite{BelBarkai06a} exhibit this behavior. A natural consequence of it 
is the non-existence of a finite correlation time, reflected by a power law
decay of the auto-correlation function, which translates itself into 
a $1/f^\beta$-behavior of the power spectrum. This type of spectrum
has been derived for a two state process of this type by Margolin and Barkai \cite[Eq.~(33)]{MargolinBarkai06}.

As said, the crucial aspect of systems with weak ergodicity breaking is that
a time average in the limit of an infinite trajectory does not
converge to a sharp value but is a random variable with a distribution of
finite width. This should also hold for the power spectrum. Therefore, 
it is not evident that an estimation of $\beta$ from a numerically computed 
 power spectrum of a single trajectory is reliable. We therefore investigate
the estimation of the power spectrum from single realizations of a 
specific process with weak ergodicity breaking.

A relevant remark is necessary: If one estimates the power spectrum by 
a discrete Fourier transform from a time series with sampling interval 
$\Delta t$ (i.e., uses the periodogram), then the variance of 
the estimator $f_k$ for the power contained in the $k$th discrete frequency is
$f_k^2$. However, if the correlations in the time domain decay exponentially
fast, then the errors in adjacent frequencies $f_k$ and $f_{k\pm 1}$ are
  sufficiently weakly correlated such that the error of a {\sl binned}
  spectrum where one takes averages over $M$ adjacent frequencies decays 
like $1/\sqrt{M}$ \cite[Sect.13.4]{PressTeukolskyVetterlingEtAl86}.
In the limit of an infinite time series, the
estimation error of the binned power spectrum of an ergodic process therefore 
decays to zero, i.e., the spectrum assumes uniquely defined values.
In the following, we will 
discuss results for binned power spectra of processes with weak ergodicity
breaking. The binning implies that 
in the analytical calculations we are allowed to 
ignore the time discretization which is present in every numerical data
analysis.

The next Sections are devoted to an analytical derivation of the 
fact that the binned power spectrum of a weakly non-ergodic process 
is a random variable whose distribution has a non-zero variance. 
In addition, we compute the correlations between different bins. 
We will show that because of these correlations, further binning cannot 
reduce the uncertainty about the true power, and moreover, that 
the whole uncertainty about the power spectrum reduces to 
a random normalization factor. This has the important consequence that
the power law exponent $\beta$ can indeed be estimated numerically regardless 
of the difficulties to estimate the spectrum itself. 

In Section \ref{sectionNumericalSimulations} we illustrate the analytical (asymptotic) results 
by numerical simulations, also opposing time averages to ensemble averages.
Note that because of the long range correlations, cutting 
a long trajectory into pieces and treat these as an ensemble is 
not valid, since this ensemble would not represent an independent sample of 
initial conditions.

\section{The model and basic approach}

Rebenshtok and Barkai introduced the following model for a thermodynamic
system showing weak ergodicity breaking \cite{RebenshtokBarkai07,RebenshtokBarkai08}:
In its simplest form it is characterized by two distributions, one waiting time
distribution with density $\phi(t)$ and one distribution of an observable
$x \in \mathbb{R}$ according to the probability density $\kappa(x)$.
In the model let $\chi_0, \chi_1, \dotsc$ be i.i.d. random variables
distributed according to $\kappa(x)$ and let $\tau_0,\tau_1,\dotsc$ be i.i.d. waiting times
distributed according to $\phi(t)$. The process $X(t)$ takes the value
$\chi_0$ in the time $0\leq t < \tau_0$ and $\chi_1$ for the next time interval of length $\tau_1$.
In general
\begin{equation}
X(t)= \chi_i \quad \text{ for } T_{i-1} \leq t < T_i
\end{equation}
with
\begin{equation}
T_i = \sum_{j=0}^{i-1} \tau_j.
\end{equation}

We assume that the first four moments of $\kappa(x)$ are
finite, i.e.,
\begin{equation}
\mu_i = \int \dop x \, x^i \kappa(x) < \infty \quad \text{for } i=1,2,3,4.
\label{definitionMoments}
\end{equation}
Moreover, we assume that the distribution is centered, i.e., $\mu_1 = 0$.
The waiting time distribution $\phi(t)$ should be in the domain of normal attraction
of an one-sided L\'{e}vy stable distribution with exponent $\alpha$ ($0 < \alpha < 1$). 
Therefore, the Laplace transform $\hat{\phi}(\lambda)$ of $\phi(t)$,
\begin{equation}
\hat{\phi}(\lambda) = \int \dop t \, e^{-\lambda t} \phi(t),
\end{equation}
can be expanded as
\begin{equation}
\hat{\phi}(\lambda) = 1 - ( \timeconstant \lambda )^\alpha + o(\lambda^\alpha) \quad \text{as } \lambda \to 0+
\label{expansionLaplaceWaitingTime}
\end{equation}
with the scaling parameter $\timeconstant>0$.

The Fourier transform of the time series $X(t)$ up to time $T$ is given by
\begin{equation}
F_T(\omega) = \int_0^T \dop t \, e^{i\omega t} X(t).
\end{equation}
The spectrum may be estimated from
\begin{equation}
S_T(\omega) = \frac{1}{n(T)} F_T(\omega) F_T(-\omega).
\label{spectralEstimateSimple}
\end{equation}
The function $n(T)$ denotes a normalization. Normally, it takes the form
$n(T)=T$, but this model requires to take $n(T)= T^\alpha$ for a non trivial
spectrum. For finite time series one would have to add correction terms for
the fact that the basis for Eq.~\eqref{spectralEstimateSimple} is a biased
estimate of the correlation function (e.g., see \cite[chapter 8.1]{KammeyerKroschel89}).
These terms vanish for the asymptotic case $T \to \infty$ which
we are considering here. If we have an ensemble of realizations, one can look
at the unbinned spectrum
\begin{equation}
S_{\text{ub}}(\omega) = \lim_{T \to \infty} S_T(\omega).
\end{equation}
Denoting the ensemble average by $\langle \cdot \rangle$, we are interested
in the expectation value of the spectrum
\begin{equation}
\langle S_{\text{ub}}(\omega) \rangle = \lim_{T \to \infty} \frac{1}{T^\alpha} \langle F_T(\omega) F_T(-\omega) \rangle
\label{unbinnedExpectation}
\end{equation}
and the covariances
\begin{equation}
\begin{split}
\langle &S_{\text{ub}}(\omega_1) S_{\text{ub}}(\omega_2) \rangle \\
&= \lim_{T \to \infty} \frac{1}{T^{2\alpha}} \langle F_T(\omega_1) F_T(-\omega_1) F_T(\omega_2) F_T(-\omega_2) \rangle.
\end{split}
\label{unbinnedCovariance}
\end{equation}
The limits Eqs.~\eqref{unbinnedExpectation} and \eqref{unbinnedCovariance} can
be rewritten to
\begin{equation}
\begin{split}
\langle &S_{\text{ub}}(\omega) \rangle
= \biggl. \lim_{r \to \infty} \frac{1}{r^\alpha} \langle F_{rT_1}(\omega) F_{rT_2}(-\omega) \rangle \biggr|_{T_1=T_2=1}, \\
\langle &S_{\text{ub}}(\omega_1) S_{\text{ub}}(\omega_2) \rangle \\
&= \biggl. \lim_{r \to \infty} \frac{1}{r^{2\alpha}} \begin{aligned}[t] \langle &F_{rT_1}(\omega_1) F_{rT_2}(-\omega_1) 
  F_{rT_3}(\omega_2) F_{rT_4}(-\omega_2) \rangle 
   \biggr|_{\substack{T_1=T_2=1 \\ T_3=T_4=1}}. \end{aligned}
\end{split}
\label{unbinnedLimitMulti}
\end{equation}
In the following, it is helpful to define the two- and four point correlations
\begin{equation}
\begin{split}
&C_2(t_1,t_2) = \langle X(t_1) X(t_2) \rangle, \\
&C_4(t_1,t_2,t_3,t_4) = \langle X(t_1) X(t_2) X(t_3) X(t_4) \rangle
\end{split}
\end{equation}
and their (double resp. quadruple) Laplace transforms
\begin{equation}
\begin{split}
\hat{C}_2(\lambda_1,\lambda_2) &= \Laplacetransform \left[ C_2(t_1,t_2) \right] \\
                 &= \int \dop t_1 \, \int \dop t_2 \, e^{-\lambda_1 t_1 -\lambda_2 t_2} C_2(t_1,t_2), 
\end{split}
\end{equation}
\begin{equation}
\begin{split}
\hat{C}_4(\lambda_1,\lambda_2,\lambda_3,\lambda_4) 
&= \Laplacetransform \left[ C_4(t_1,t_2,t_3,t_4) \right] \\
&= \int \dop^4 t \, e^{-\bl \mathbf{t}} C_4(\mathbf{t}).
\end{split}
\end{equation}
The double resp. quadruple Laplace transforms of the expressions in Eq.~\eqref{unbinnedLimitMulti}
are
\begin{equation}
\begin{split}
\Laplacetransform &\bigl[ \langle F_{T_1}(\omega) F_{T_2}(-\omega) \rangle \bigr]
= \frac{1}{\lambda_1 \lambda_2} \hat{C}_2(\lambda_1 - i \omega, \lambda_2 + i\omega), \\
\Laplacetransform &\bigl[ \langle F_{T_1}(\omega_1) F_{T_2}(-\omega_1) F_{T_3}(\omega_2) F_{T_4}(-\omega_2) \rangle \bigr] \\
&= \frac{1}{\lambda_1 \lambda_2 \lambda_3 \lambda_4}
   \hat{C}_4(\lambda_1 - i \omega_1, \lambda_2 + i \omega_1, \lambda_3 - i \omega_2, \lambda_4 + i \omega_2).
\end{split}
\end{equation}
The limits in Eq.~\eqref{unbinnedLimitMulti} can also be expressed in
Laplace space using a formulation following the multidimensional Tauberian theorem by Drozhzhinov and Zav'jalov
\cite{DrozhzhinovZavjalov80,Drozhzhinov83} as
\begin{align}
\begin{split}
\Laplacetransform &\Bigl[ \lim_{r \to \infty} \frac{1}{r^\alpha} \langle F_{rT_1}(\omega) F_{rT_2}(-\omega) \rangle \Bigr] \\
&= \lim_{\zeta \to 0+} \frac{\zeta^\alpha}{\lambda_1 \lambda_2}
   \hat{C}_2(\zeta \lambda_1 - i \omega, \zeta\lambda_2 + i\omega),  
\end{split} \label{limitUnbinnedExpectation} \displaybreak[0] \\
\begin{split}
\Laplacetransform &\Bigl[ \lim_{r \to \infty} \frac{1}{r^{2\alpha}} \langle F_{rT_1}(\omega_1) F_{rT_2}(-\omega_1) 
F_{rT_3}(\omega_2) F_{rT_4}(-\omega_2) \rangle \Bigr] \\
&= \lim_{\zeta \to 0+} \frac{\zeta^{2\alpha}}{\lambda_1 \lambda_2 \lambda_3 \lambda_4}
   \hat{C}_4(\begin{aligned}[t] \zeta&\lambda_1 - i \omega_1, \zeta\lambda_2 + i \omega_1, \\
                                &\zeta\lambda_3 - i \omega_2, \zeta\lambda_4 + i \omega_2). \end{aligned}
\end{split} \label{limitUnbinnedCovariance}
\end{align}
In the next section, we will show how one can calculate the Laplace transforms
$\hat{C}_n(\bl)$ of the multi point correlations $C_n(\mathbf{t})$.

However, when only one time series is given,
taking $S_T(\omega)$ from Eq.~\eqref{spectralEstimateSimple} as estimate
of the spectrum will lead to unsatisfactory results as the value of $S_T(\omega)$
will fluctuate except for some pathological cases. One often relies to the concept
of binning. Instead of considering a single frequency $\omega$, one averages all
available frequencies in an given interval 
$[\omega - \frac{1}{2} \Delta \omega, \omega + \frac{1}{2} \Delta \omega]$.
For a time series of length $T$ the Fourier transform returns values at
frequencies which are evenly spaced with distance $\frac{2\pi}{T}$. For large $T$ the
sum can be approximated by an integral and
we can consider as observable for the binned spectrum
\begin{equation}
S_T(\omega,\Delta \omega) = \frac{1}{\Delta \omega} 
\int_{\omega - \frac{1}{2} \Delta \omega}^{\omega + \frac{1}{2} \Delta \omega}
\dop \omega^\prime \, S_T(\omega^\prime).
\label{spectralEstimateBinning}
\end{equation}
The spectrum can be estimated from
\begin{equation}
S_{\text{bin}}(\omega,\Delta \omega) = \lim_{T\to \infty} S_T(\omega,\Delta \omega).
\end{equation}
Analytically, it is helpful to let the bin size go to zero as a last step
\begin{equation}
S_{\text{bin}}(\omega) = \lim_{\Delta \omega \to 0} S(\omega, \Delta \omega).
\end{equation}
It is important to take the last limit at the very end. In most of the important cases, the
value of $S_{\text{bin}}(\omega)$ will converge almost surely to the
value of the spectrum.

In this paper, we want to consider the situation where the stochastic
process $X(t)$ is the model by Rebenshtok and Barkai. The estimation of
the spectrum is connected with averaging over the time series, 
therefore the question arises
how the weak ergodicity breaking affects the observable $S_{\text{bin}}(\omega)$.
Does the value of $S_{\text{bin}}(\omega)$ converge to a single value or does it
converge to a non trivial probability distribution? It will turn out
that the latter is the case. Unlike the observable discussed by
Rebenshtok and Barkai \cite{RebenshtokBarkai07,RebenshtokBarkai08},
it seems not to be possible to determine directly the probability
distribution of $S_{\text{bin}}(\omega)$. But one can obtain important
information from the moments $\langle S_{\text{bin}}(\omega) \rangle$ and
$\langle S_{\text{bin}}(\omega_1) S_{\text{bin}}(\omega_2) \rangle$.

The first step is to look at $\langle S(\omega,\Delta \omega) \rangle$
with $2|\omega| > \Delta \omega$:
\begin{equation}
\langle S_{\text{bin}}(\omega,\Delta \omega) \rangle
= \lim_{T \to \infty} 
\frac{T^{-\alpha}}{\Delta \omega} \int_{\omega - \frac{1}{2} \Delta \omega}^{\omega + \frac{1}{2} \Delta \omega}
\dop \omega' \, \langle F_T(\omega') F_T(-\omega') \rangle.
\end{equation}
We proceed similarly to the unbinned case and formulate the limit in
Laplace space (note that the Laplace transform and the $\omega$-integral
commute by Fubini's theorem and the estimate 
$|\langle F_{T_1}(\omega) F_{T_2}(-\omega) \rangle | \leq \mu_2 T_1 T_2$
where $\mu_2$ is defined in Eq.~\eqref{definitionMoments})
\begin{equation}
\begin{split}
\Laplacetransform&\left[ \lim_{r \to \infty} r^{-\alpha}
\frac{1}{\Delta \omega} \int_{\omega - \frac{1}{2} \Delta \omega}^{\omega + \frac{1}{2} \Delta \omega} \dop\omega' \,
\langle F_{r T_1}(\omega') F_{rT_2}(-\omega') \rangle \right] \\
&\! = \lim_{\zeta \to 0+} \frac{\zeta^\alpha}{\Delta \omega} \int_{\omega - \frac{1}{2} \Delta \omega}^{\omega + \frac{1}{2} \Delta \omega}
\dop \omega' \,
\frac{1}{\lambda_1 \lambda_2} \hat{C}_2(\zeta \lambda_1 - i \omega', \zeta \lambda_2 + i \omega').
\end{split}
\label{LaplaceTwoPointLimes}
\end{equation}

Similarly, $\langle S(\omega_1,\Delta \omega_1) S(\omega_2,\Delta \omega_2) \rangle$
with $2|\omega_1| > \Delta \omega_1$ and $2|\omega_2| > \Delta \omega_2$ can
be obtained from
\begin{equation}
\begin{split}
\langle &S_{\text{bin}}(\omega_1,\Delta \omega_1) S_{\text{bin}}(\omega_2,\Delta \omega_2) \rangle \\
&= \begin{aligned}[t] \biggl. \lim_{r \to \infty}& r^{-2\alpha}
\frac{1}{\Delta \omega_1 \Delta \omega_2} 
\int_{\omega_1 - \frac{1}{2} \Delta \omega_1}^{\omega_1 + \frac{1}{2} \Delta \omega_1} \dop \omega_1' \,
\int_{\omega_2 - \frac{1}{2} \Delta \omega_2}^{\omega_2 + \frac{1}{2} \Delta \omega_2} \dop \omega_2' \\
&\langle F_{r T_1}(\omega_1') F_{rT_2}(-\omega_1') F_{r T_3}(\omega_2') F_{rT_4}(-\omega_2') \rangle 
  \biggl|_{\substack{T_1=T_2=1 \\ T_3=T_4=1}}. \end{aligned}
\end{split}
\end{equation}
One gets
\begin{equation}
\begin{split}
\Laplacetransform &\biggl[\begin{aligned}[t] 
 \lim_{r \to \infty} \frac{r^{-2\alpha}}{\Delta \omega_1 \Delta \omega_2}
 &\int_{\omega_1 - \frac{1}{2} \Delta \omega_1}^{\omega_1 + \frac{1}{2} \Delta \omega_1} \dop \omega_1' \,
\int_{\omega_2 - \frac{1}{2} \Delta \omega_2}^{\omega_2 + \frac{1}{2} \Delta \omega_2} \dop \omega_2' \\
&\langle F_{r T_1}(\omega_1') F_{rT_2}(-\omega_1') F_{r T_3}(\omega_2') F_{rT_4}(-\omega_2') \rangle
\biggr] \end{aligned} \\
&= \begin{aligned}[t] \lim_{\zeta \to 0+} &\frac{\zeta^{2\alpha}}{\Delta \omega_1 \Delta \omega_2}
\int_{\omega_1 - \frac{1}{2} \Delta \omega_1}^{\omega_1 + \frac{1}{2} \Delta \omega_1} \dop \omega_1' \,
\int_{\omega_2 - \frac{1}{2} \Delta \omega_2}^{\omega_2 + \frac{1}{2} \Delta \omega_2} \dop \omega_2' 
\frac{1}{\lambda_1 \lambda_2 \lambda_3 \lambda_4} \\
&\hat{C}_4(\zeta \lambda_1 - i \omega_1',\zeta  \lambda_2 + i \omega_1',\zeta  \lambda_3 - i \omega_2',\zeta  \lambda_4 + i \omega_2').
\end{aligned}
\end{split}
\label{fourPointLimes}
\end{equation}
Therefore the problem can be reduced to the determination of the Laplace transforms $\hat{C}_n(\bl)$
of the $n$ point correlations functions. We describe in the next section how to obtain them.

\section{A diagrammatic approach to the multi point correlation functions}

In \cite{NiemannKantz08} a diagrammatic method was introduced how to determine the joint probability
distributions and multi point correlation functions of a continuous-time random walk.
In this section we are adapting this method to determine $\hat{C}_n(\bl)$.
For any set of natural numbers $q_1,\dotsc,q_n \in \mathbb{N}_0$, one defines
\begin{equation}
C_n[\mathbf{q}](\mathbf{t})
= \left\langle \prod_{k=1}^n \chi_{q_k} 1_{[T_{q_k},T_{q_k+1}[}(t_k) \right\rangle
\end{equation}
with the indicator function ($A$ being any subset $A \subseteq \mathbb{R}$)
\begin{equation}
1_A(t) = \begin{cases} 1 & \text{for } t \in A, \\
0 & \text{for } t \notin A.
\end{cases}
\end{equation}
The $C_n[\mathbf{q}](\mathbf{t})$ can be considered as partial correlations which
only contribute, when the time $t_i$ is in the $(q_i+1)$th waiting time (for $i=1,\dotsc,n$).
Therefore,
\begin{equation}
\begin{split}
C_n(\mathbf{t}) &= \langle X(t_1) \dotsm X(t_n) \rangle \\
&= \sum_{q_1,\dotsc,q_n=0}^\infty C_n[\mathbf{q}](\mathbf{t}).
\end{split}
\end{equation}
Define the contribution of one step as
\begin{equation}
\eta^{(j)}_n[\mathbf{q}](\mathbf{t})
= \prod_{i:q_i=j} \chi_j 1_{[0,\tau_j[}(t_i) \prod_{i:q_i>j} \delta(t_i - \tau_j) \prod_{i:q_i<j} \delta(t_i).
\end{equation}
One obtains by induction
\begin{equation}
C_n[\mathbf{q}](\mathbf{t}) = \left\langle \underset{j}{\bigstar} \eta^{(j)}_n[\mathbf{q}](\mathbf{t}) \right\rangle
\end{equation}
where $\star$ denotes the convolution in the time variable $t_1, \dotsc, t_n$ and $j$ runs
from $0$ to any natural number larger than $\max(q_1,\dotsc,q_n)$.

Transforming into Laplace space gives
\begin{equation}
\begin{split}
\hat{C}_n[\mathbf{q}](\bl) &= \Laplacetransform\left[ C_n[\mathbf{q}](\mathbf{t})\right] \\
&= \prod_j \langle \hat{\eta}^{(j)}_n [\mathbf{q}] (\bl) \rangle.
\end{split}
\end{equation}
Using the following notations from \cite{NiemannKantz08} (suppressing the dependence on $\mathbf{q}$)
\begin{equation}
\singlestepset{V}_j = \{i : q_j = i \}, \quad 
\singlestepset{E}_j = \{i:q_i<j\}, \quad 
\singlestepset{L}_j = \{i:q_i>j\}. 
\end{equation}
The set $\singlestepset{V}_j$ (``vertex'') contains the indices which are 
at position $j$, the set $\singlestepset{E}_j$ (``earlier'') contains the indices
which are before that position and the set $\singlestepset{L}_j$ (``later'')
contains the indices which are after that position.
With the notation
\begin{equation}
\Lambda_\setJ = \sum_{j \in \setJ} \lambda_j
\end{equation}
one gets
\begin{equation}
\begin{split}
\langle \hat{\eta}^{(j)}_n [\mathbf{q}] (\bl) \rangle
&= \biggl\langle \int \mathrm{d}^n t \, e^{- \bl \mathbf{t}} 
      \prod_{i \in \singlestepset{V}_j} \chi_j 1_{[0,\tau_j[}(t_i) 
      \prod_{i \in \singlestepset{E}_j} \delta(t_i - \tau_j) 
      \prod_{i \in \singlestepset{L}_j} \delta(t_i) 
   \biggr\rangle \\
&= \frac{\mu_{|\singlestepset{V}_j|}}{\prod_{v \in \singlestepset{V}_j} \lambda_v} 
\sum_{\setJ \in \powerset(\singlestepset{V}_j)} (-1)^{|\setJ|} \hat{\phi}(\Lambda_{\setJ\cup \singlestepset{L}_j}). \\
\end{split}
\end{equation}
Here, the power set of $\singlestepset{V}_j$ is denoted by $\powerset(\singlestepset{V}_j)$ and
$|\setJ|$ is the number of elements of $\setJ$. Therefore, the last sum goes over all
subsets $\setJ$ of indices of $\singlestepset{V}_j$.

\begin{figure}
\includegraphics[width=85mm]{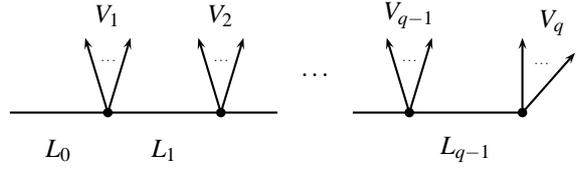}
\caption{A generic diagram}
\label{genericDiagram}
\end{figure}

As in \cite{NiemannKantz08} we introduce diagrams which denote the
relative ordering of the $q_i$s.
A generic diagram is given in Fig.~\ref{genericDiagram}.
The main idea behind this consists in grouping all steps $i$
with the same $\singlestepset{V}_i$, $\singlestepset{E}_i$ and $\singlestepset{L}_i$.
If $\singlestepset{V}_i=\{\}$ then there can be several successive steps with
the same $\singlestepset{E}_i$ and $\singlestepset{L}_i$. These successive steps
are combined to a horizontal line. If $\singlestepset{V}_i\neq\{\}$ then there
can be only one step with these sets of indices. This is represented by a vertex
with the indices $\singlestepset{V}_i$ leaving. Therefore the diagram Fig.~\ref{genericDiagram}
represents all $\mathbf{q}$ with the following properties: First there is any
number of steps $i$ with $\singlestepset{E}_i = \{\}$, $\singlestepset{V}_i=\{\}$ and
$\singlestepset{L}_i = \diagramset{L}_0$. A single step with $\singlestepset{E}_i=\{\}$,
$\singlestepset{V}_i = \diagramset{V}_1$ and $\singlestepset{L}_i=\diagramset{L}_1$ follows.
At next is any number of steps $i$ with $\singlestepset{E}_i = \diagramset{V}_1$,
$\singlestepset{V}_i=\{\}$ and $\singlestepset{L}_i=\diagramset{L}_1$. The rest in
interpreted analogously. For each $n$ point correlation there is only a finite
number of diagrams and summing over all $\mathbf{q}$ represented by a diagram
turns out to be easy. It turns out that one can associate to each part of
the diagram a factor (for a more comprehensive derivation of these results,
we refer to \cite{NiemannKantz08}).
One obtains the following rules (where $\mu_i$ is defined in Eq.~\eqref{definitionMoments})
\begin{equation}
\begin{split}
&\gamma_{\text{vertex } i}=\frac{\mu_{|\diagramset{V}_i|}}{\prod_{v \in \diagramset{V}_i \lambda_v}}
\sum_{\setJ \in \powerset(\diagramset{V}_i)} (-1)^{|\setJ|} \hat{\phi}(\Lambda_{\setJ\cup \diagramset{L}_i}), \\
&\gamma_{\text{line } i}= \sum_{j=0}^\infty \hat{\phi}^j (\Lambda_{\diagramset{L}_i}) 
=\frac{1}{1 - \hat{\phi}(\Lambda_{\diagramset{L}_i})}.
\end{split}
\end{equation}

\begin{figure}
\includegraphics[width=85mm]{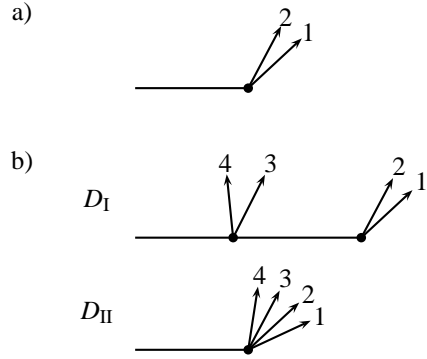}
\caption{The types of diagrams relevant for a) the two point correlation and b) four point correlation}
\label{diagramsCorrelations}
\end{figure}

Since we assumed that $\mu_1 = 0$, we do not need to consider diagrams which
contain vertices with only one leaving line. Therefore a single diagram
remains for $\hat{C}_2(\lambda_1,\lambda_2)$ which drawn in Fig.~\ref{diagramsCorrelations}a. This
gives
\begin{equation}
\begin{split}
\hat{C}_2(\lambda_1,\lambda_2) &= \gamma_{\text{line}, \diagramset{L}=\{1,2\}} \gamma_{\text{vertex}, \diagramset{V}=\{1,2\},\diagramset{L}=\{\}} \\
&= \frac{\mu_2}{\lambda_1 \lambda_2} \frac{1 - \hat{\phi}(\lambda_1) - \hat{\phi}(\lambda_2) + \hat{\phi}(\lambda_1+\lambda_2)}{1-\hat{\phi}(\lambda_1+\lambda_2)}.
\end{split}
\label{resultTwoPointCorrelation}
\end{equation}

For the four point correlation two types of diagrams are relevant, which are
drawn in Fig.~\ref{diagramsCorrelations}b. The contributions are
\begin{equation}
\begin{split}
&D_{\text{I}}(\bl) \begin{aligned}[t]
&= \gamma_{\text{line}, \diagramset{L}=\{1,2,3,4\}} \gamma_{\text{vertex}, \diagramset{V}=\{3,4\},\diagramset{L}=\{1,2\}} \\
&\phantom{=} \times \gamma_{\text{line}, \diagramset{L}=\{1,2\}} \gamma_{\text{vertex}, \diagramset{V}=\{1,2\},\diagramset{L}=\{\}}
\end{aligned} \\
&= \frac{\mu_2^2}{\lambda_1 \lambda_2 \lambda_3 \lambda_4} 
  \frac{\hat{\phi}(\Lambda_{\{1,2\}}) - \hat{\phi}(\Lambda_{\{1,2,3\}}) - \hat{\phi}(\Lambda_{\{1,2,4\}}) + \hat{\phi}(\Lambda_{\{1,2,3,4\}})}
     {1 - \hat{\phi}(\Lambda_{\{1,2,3,4\}})} \\
&\phantom{=} \times \frac{1 - \hat{\phi}(\lambda_1) - \hat{\phi}(\lambda_2) + \hat{\phi}(\lambda_1+\lambda_2)}
     {1 - \hat{\phi}(\lambda_1+\lambda_2)} 
\end{split} \label{resultD1} 
\end{equation}
\begin{equation}
\begin{split}
D_{\text{II}}(\bl)
=& \gamma_{\text{line}, \diagramset{L}=\{1,2,3,4\}} \gamma_{\text{vertex}, \diagramset{V}=\{1,2,3,4\},\diagramset{L}=\{\}} \\
=& \frac{\mu_4}{\lambda_1 \lambda_2 \lambda_3 \lambda_4} \frac{\sum_{\diagramset{J} \in \diagramset{P}(\{1,2,3,4\})} 
(-1)^{|\diagramset{J}|} \hat{\phi}(\Lambda_\diagramset{J})}
{1 - \hat{\phi}(\Lambda_{\{1,2,3,4\}})}.
\end{split} \label{resultD2}
\end{equation}
The function $\hat{C}_4(\bl)$ is obtained by summing over all possible
combinations of indices at the vertices, i.e.,
\begin{equation}
\begin{split}
\hat{C}_4(\bl)
=& D_{\text{I}}(\lambda_1,\lambda_2,\lambda_3,\lambda_4) + D_{\text{I}}(\lambda_1,\lambda_3,\lambda_2,\lambda_4) \\
&+ D_{\text{I}}(\lambda_1,\lambda_4,\lambda_2,\lambda_3) + D_{\text{I}}(\lambda_2,\lambda_3,\lambda_1,\lambda_4) \\
&+ D_{\text{I}}(\lambda_2,\lambda_4,\lambda_1,\lambda_3) + D_{\text{I}}(\lambda_3,\lambda_4,\lambda_1,\lambda_2) \\
&+D_{\text{II}}(\lambda_1,\lambda_2,\lambda_3,\lambda_4).
\end{split}
\label{resultFourPointCorrelation}
\end{equation}

\section{Evaluating the spectral observables}

\subsection{The unbinned observables}

We can use the results of the last section to determine the spectral
observables. Using the expansion Eq.~\eqref{expansionLaplaceWaitingTime}
and the two point correlations Eq.~\eqref{resultTwoPointCorrelation}
to calculate the limit Eq.~\eqref{limitUnbinnedExpectation}
\begin{equation}
\begin{split}
&\lim_{\zeta \to 0+} \frac{\zeta^\alpha}{\lambda_1 \lambda_2}
   \hat{C}_2(\zeta \lambda_1 - i \omega, \zeta\lambda_2 + i\omega) \\
&\quad = \frac{1}{\lambda_1\lambda_2 (\lambda_1 + \lambda_2)^\alpha}
  \frac{\mu_2}{\timeconstant^\alpha} \frac{2 - \hat{\phi}(i\omega) - \hat{\phi}(-i\omega)}{\omega^2}
\end{split} 
\end{equation}
The Laplace transform can be inverted by using
\begin{equation}
\Laplacetransform \left[ \frac{\min(T_1,T_2)^\alpha}{\Gamma(1+\alpha)} \right] = \frac{1}{\lambda_1 \lambda_2(\lambda_1+\lambda_2)^\alpha}.
\end{equation}
Therefore, one has
\begin{equation}
\langle S_{\text{ub}}(\omega) \rangle
= \frac{\mu_2}{\timeconstant^\alpha \Gamma(1+\alpha)} \frac{2 - \hat{\phi}(i\omega) - \hat{\phi}(-i\omega)}{\omega^2} 
\label{expectationUnbinnedSpectrum}
\end{equation}
with the behavior near $\omega \to 0+$
\begin{equation}
\langle S_{\text{ub}}(\omega) \rangle
= \frac{2 \mu_2 \cos\left( \frac{\pi}{2} \alpha \right)}{\timeconstant^\alpha \Gamma(1+\alpha)}
    \frac{1}{|\omega|^{2-\alpha}} + o \left( \frac{1}{|\omega|^{2-\alpha}} \right).
\label{expectationSpectrumOrigin}
\end{equation}
This last equation shows that the spectrum has a $1/f^{2-\alpha}$ form near the origin.

Following \cite[Lemma XV.1.3]{Feller71}, there exists an $\Omega \geq 0$
such that
\begin{equation}
\hat{\phi}(i\omega) = 1 \quad \Leftrightarrow \quad \omega \in \Omega \mathbb{Z}.
\label{explanationOmega}
\end{equation}
The case $\Omega \neq 0$ appears if and only if the waiting times $\tau_i$ can take
only the values $0, \frac{2\pi}{\Omega}, \frac{4\pi}{\Omega}, \dotsc$. Therefore,
more generally,
\begin{equation}
\hat{\phi}(z) = \hat{\phi}(z + i\omega) \quad \text{for } 
\operatorname{Re}(z) \geq 0 \text{ and } \omega \in \Omega \mathbb{Z}.
\end{equation}
In this paper we concentrate on the spectrum for frequencies $\omega \notin \Omega \mathbb{Z}$.
The frequencies $\omega \in \Omega \mathbb{Z}$ behave differently, e.g., the behavior of
$S_{\mathrm{ub}}(0)$ is completely described by the result of Rebenshtok and Barkai \cite{RebenshtokBarkai07, RebenshtokBarkai08}.
Now, we can proceed to calculate the limit Eq.~\eqref{limitUnbinnedCovariance} by looking at the
terms Eqs.~\eqref{resultD1} and \eqref{resultD2} with the parameters given in Eq.~\eqref{resultFourPointCorrelation}.
We have
\begin{equation}
\begin{split}
&\lim_{\zeta \to 0+} \zeta^{2\alpha} \begin{aligned}[t] 
\Bigl( &
D_{\text{I}}(\zeta \lambda_1 - i \omega_1,\zeta  \lambda_2 + i \omega_1,\zeta  \lambda_3 - i \omega_2,\zeta  \lambda_4 + i \omega_2) \\
&+ D_{\text{I}}(\zeta  \lambda_3 - i \omega_2,\zeta  \lambda_4 + i \omega_2,\zeta \lambda_1 - i \omega_1,\zeta  \lambda_2 + i \omega_1)
\Bigr) \end{aligned} \\
&= \Gamma(1+\alpha)^2
  \frac{1}{\Lambda_{\{1,2,3,4\}}^\alpha} \left( \frac{1}{\Lambda_{\{1,2\}}^\alpha} + \frac{1}{\Lambda_{\{3,4\}}^\alpha} \right)  
 \langle S_{\text{ub}}(\omega_1) \rangle \langle S_{\text{ub}}(\omega_2) \rangle.
\end{split}
\end{equation}
Using the fact $\hat{\phi}(i\omega_1) 1_{\Omega \mathbb{Z}}(\omega_1 - \omega_2) = \hat{\phi}(i\omega_2) 1_{\Omega \mathbb{Z}}(\omega_1 - \omega_2)$,
yields similarly
\begin{equation}
\begin{split}
&\lim_{\zeta \to 0+} \zeta^{2\alpha} \begin{aligned}[t]
\Bigl( &
D_{\text{I}}(\zeta \lambda_1 - i \omega_1,\zeta  \lambda_4 + i \omega_2,\zeta  \lambda_2 + i \omega_1,\zeta  \lambda_3 - i \omega_2) \\
&+ D_{\text{I}}(\zeta  \lambda_2 + i \omega_1,\zeta  \lambda_3 - i \omega_2, \zeta \lambda_1 - i \omega_1,\zeta  \lambda_4 + i \omega_2)
\Bigr) \end{aligned} \\
&= \Gamma(1+\alpha)^2
  \frac{1}{\Lambda_{\{1,2,3,4\}}^\alpha} \left( \frac{1}{\Lambda_{\{2,3\}}^\alpha} + \frac{1}{\Lambda_{\{1,4\}}^\alpha} \right)  
  \langle S_{\text{ub}}(\omega_1) \rangle \langle S_{\text{ub}}(\omega_2) \rangle \\
&\phantom{=}\times 1_{\Omega \mathbb{Z}}(\omega_1 - \omega_2)
\end{split}
\end{equation}
and
\begin{equation}
\begin{split}
&\lim_{\zeta \to 0+} \zeta^{2\alpha} \begin{aligned}[t]
\Bigl( &
D_{\text{I}}(\zeta \lambda_1 - i \omega_1,\zeta  \lambda_3 - i \omega_2,\zeta  \lambda_2 + i \omega_1,\zeta  \lambda_4 + i \omega_2) \\
&+ D_{\text{I}}(\zeta  \lambda_2 + i \omega_1,\zeta  \lambda_4 + i \omega_2, \zeta \lambda_1 - i \omega_1,\zeta  \lambda_3 - i \omega_2)
\Bigr) \end{aligned} \\
&= \Gamma(1+\alpha)^2
  \frac{1}{\Lambda_{\{1,2,3,4\}}^\alpha} \left( \frac{1}{\Lambda_{\{1,3\}}^\alpha} + \frac{1}{\Lambda_{\{2,4\}}^\alpha} \right) 
  \langle S_{\text{ub}}(\omega_1) \rangle \langle S_{\text{ub}}(\omega_2) \rangle \\
&\phantom{=}\times 1_{\Omega \mathbb{Z}}(\omega_1 + \omega_2).
\end{split}
\end{equation}
Finally
\begin{equation}
\lim_{\zeta \to 0+} \zeta^{2\alpha} 
D_{\text{II}}(\zeta \lambda_1 - i \omega_1,\zeta  \lambda_4 + i \omega_2,\zeta  \lambda_2 + i \omega_1,\zeta  \lambda_3 - i \omega_2)
=0.
\end{equation}
In \ref{appendixLaplace}, we derive the following Laplace transform
\begin{equation}
\begin{split}
\Laplacetransform &\begin{aligned}[t] \biggl[ \min(T_1,&T_2,T_3,T_4)^\alpha \min(T_1,T_2)^\alpha  \\
                     &\times F(\alpha,-\alpha; 1+\alpha; \frac{\min(T_1,T_2,T_3,T_4)}{\min(T_1,T_2)} ) \biggr] \end{aligned} \\
&= \frac{\Gamma(1+\alpha)^2}{\lambda_1\lambda_2\lambda_3\lambda_4} \frac{1}{\Lambda_{\{1,2,3,4\}}^\alpha} \frac{1}{\Lambda_{\{1,2\}}^\alpha}
\end{split}
\label{LaplaceTransformFourTimes}
\end{equation}
with the hypergeometric function $F(\alpha,-\alpha;1+\alpha;x)$.
As
\begin{equation}
F(\alpha,-\alpha;1+\alpha;1) = \frac{\Gamma(1+\alpha)^2}{\Gamma(1+2\alpha)}
\end{equation}
we obtain for the limit Eq.~\eqref{unbinnedLimitMulti}
\begin{equation}
\begin{split}
\langle S_{\text{ub}}(\omega_1) S_{\text{ub}}(\omega_2) \rangle
=& 2 \frac{\Gamma(1+\alpha)^2}{\Gamma(1+2\alpha)} \langle S_{\text{ub}}(\omega_1) \rangle \langle S_{\text{ub}}(\omega_2) \rangle \\
 &\times  \left( 1 + 1_{\Omega \mathbb{Z}}(\omega_1 - \omega_2) + 1_{\Omega \mathbb{Z}}(\omega_1 + \omega_2) \right).
\end{split}
\label{unbinnedCovarianceGeneral}
\end{equation}
Eq.~\eqref{unbinnedCovarianceGeneral} will play an important role in
determining the properties of the binned case. 
In particular, the last equation 
determines the variance of $S_{\text{ub}}(\omega)$ (for $2 \omega \notin \Omega \mathbb{Z}$)
\begin{equation}
\operatorname{Var}[S_{\text{ub}}(\omega)]
= V_{\text{ub}}(\alpha) \langle S_{\text{ub}}(\omega) \rangle^2 
\label{resultUnbinnedVariance}
\end{equation}
with the function
\begin{equation}
V_{\text{ub}}(\alpha) = 4 \frac{\Gamma(1+\alpha)^2}{\Gamma(1+2\alpha)} - 1.
\label{definitionVUnbinned}
\end{equation}
As already mentioned in the introduction, even in the ergodic limit $\alpha \to 1$,
the value $V_{\text{ub}}(1) = 1$ is larger than zero and therefore the
observable $S_{\text{ub}}(\omega)$ does fluctuate.

The correlation coefficient between $S_{\text{ub}}(\omega_1)$ and $S_{\text{ub}}(\omega_2)$
(for $2 \omega_1 \notin \Omega \mathbb{Z}$ and $2 \omega_2 \notin \Omega \mathbb{Z}$; here
$\Omega$ denotes the periodicity of $\hat{\phi}(\lambda)$, see Eq.~\eqref{explanationOmega})
\begin{equation}
\rho[S_{\text{ub}}(\omega_1), S_{\text{ub}}(\omega_2)]
= \begin{cases}
1 & \text{for } \omega_1 - \omega_2 \in \Omega \mathbb{Z} \\
R(\alpha) & \text{for } \omega_1 - \omega_2 \notin \Omega \mathbb{Z}
\end{cases}
\label{casesUnbinnedCorrelation}
\end{equation}
with
\begin{equation}
R(\alpha) = \frac{2\Gamma(1+\alpha)^2 - \Gamma(1+2\alpha)}{4\Gamma(1+\alpha)^2 - \Gamma(1+2\alpha)}.
\label{resultUnbinnedCorrelation}
\end{equation}
In many cases we have $\Omega = 0$, i.e.,
\begin{equation}
\rho[S_{\text{ub}}(\omega_1), S_{\text{ub}}(\omega_2)]
= R(\alpha) \quad \text{for } |\omega_1| \neq |\omega_2|.
\end{equation}
In the ergodic limit, we have $R(1) = 0$. Therefore, the
observations $S_{\text{ub}}(\omega_1)$
and $S_{\text{ub}}(\omega_2)$ of the spectrum at two different
frequencies $\omega_1$ and $\omega_2$ are uncorrelated.

\subsection{The binned observables}

Here, we are going to use the results from the unbinned case to
evaluate the limits Eqs.~\eqref{LaplaceTwoPointLimes} and
\eqref{fourPointLimes}. The limits on the right hand sides of these
equations can be commuted with the frequency integrals describing
the binning. As this statement is not obvious, we are giving an argument.

First, we are considering $\langle S_{\text{bin}}(\omega, \Delta \omega) \rangle$
with $2 |\omega| > \Delta \omega$. Then we have the simple estimate
\begin{equation}
|\hat{C}_2(\zeta \lambda_1 - i \omega, \zeta \lambda_2 + i \omega)|
\leq \frac{1}{\omega^2} \frac{4}{1- \hat{\phi}(\zeta(\lambda_1 + \lambda_2))}.
\end{equation}
The commutativity follows by dominated convergence. Therefore, we get
\begin{equation}
\langle S_{\text{bin}}(\omega, \Delta \omega) \rangle 
= \frac{1}{\Delta \omega}
\int_{\omega - \frac{1}{2} \Delta \omega}^{\omega + \frac{1}{2} \Delta \omega}
\dop \omega^\prime \, \langle S_{\text{ub}}(\omega') \rangle
\end{equation}
and accordingly 
\begin{equation}
\langle S_{\text{bin}}(\omega) \rangle
= \langle S_{\text{ub}}(\omega) \rangle.
\end{equation}
It is not surprising that the binned estimate yields also the
correct expectation value for an ensemble of realizations.

The argument for $\langle S_{\text{bin}}(\omega_1,\Delta \omega_1)
S_{\text{bin}}(\omega_2,\Delta \omega_2)\rangle$ (with
$2 |\omega_1| > \Delta \omega_1$ and $2 |\omega_2| > \Delta \omega_2$)
needs more work. First notice that we have the following inequality
\begin{equation}
|1 - \hat{\phi}(\lambda + i \omega)| \geq 1 - |\hat{\phi}(\lambda + i\omega)|
\geq 1 - \hat{\phi}(\lambda).
\end{equation}
This gives the following estimates (replace the tilded variables by
the corresponding permutations of the $\lambda_i$s and $\omega_i$s)
\begin{align}
\begin{split}
&\bigl| D_{\text{I}}(\zeta \tilde{\lambda}_1 + i \tilde{\omega}_1, \zeta \tilde{\lambda}_2 + i \tilde{\omega}_2,
     \zeta \tilde{\lambda}_3 + i \tilde{\omega}_3, \zeta \tilde{\lambda}_4 + i \tilde{\omega}_4) \bigr| \\
&\quad \leq \frac{1}{| \tilde{\omega}_1  \tilde{\omega}_2 \tilde{\omega}_3 \tilde{\omega}_4 |}
      \frac{4}{1 - \hat{\phi}(\zeta \tilde{\Lambda}_{\{1,2,3,4\}})}
      \frac{4}{1 - \hat{\phi}(\zeta \tilde{\Lambda}_{\{1,2\}})}
\end{split} \displaybreak[0] \\
\intertext{and}
\begin{split}
&\bigl| D_{\text{II}}(\zeta \lambda_1 - i \omega_1, \zeta \lambda_2 + i \omega_1,
     \zeta \lambda_3 - i \omega_2, \zeta \lambda_4 + i \omega_2) \bigr| \\
&\quad \leq \frac{1}{\omega_1^2 \omega_2^2}
      \frac{16}{1 - \hat{\phi}(\zeta \Lambda_{\{1,2,3,4\}})}.
\end{split}
\end{align}
By dominated convergence and Eq.~\eqref{unbinnedCovarianceGeneral}, this leads to
\begin{equation}
\begin{split}
&\langle S_{\text{bin}}(\omega_1, \Delta \omega_1) S_{\text{bin}}(\omega_2, \Delta \omega_2) \rangle \\
&=  \frac{1}{\Delta \omega_1 \Delta \omega_2} \int_{\omega_1 - \frac{1}{2} \Delta \omega_1}^{\omega_1 + \frac{1}{2} \Delta \omega_1}
\dop \omega_1^\prime \,
 \int_{\omega_2 - \frac{1}{2} \Delta \omega_2}^{\omega_2 + \frac{1}{2} \Delta \omega_2}
\dop \omega_2^\prime \, \langle S_{\text{ub}}(\omega_1') S_{\text{ub}}(\omega_2') \rangle \\
&=  2 \frac{\Gamma(1+\alpha)^2}{\Gamma(1+2\alpha)}  \frac{1}{\Delta \omega_1 \Delta \omega_2}
\int 
\dop \omega_1^\prime \,
 \int 
\dop \omega_2^\prime \, \langle S_{\text{bin}}(\omega_1') \rangle \langle S_{\text{bin}}(\omega_2') \rangle
\end{split}
\end{equation}
or shorter
\begin{equation}
\langle S_{\text{bin}}(\omega_1) S_{\text{bin}}(\omega_2) \rangle
=  2 \frac{\Gamma(1+\alpha)^2}{\Gamma(1+2\alpha)} 
  \langle S_{\text{bin}}(\omega_1) \rangle \langle S_{\text{bin}}(\omega_2) \rangle
\label{binnedCovarianceGeneral}
\end{equation}
for all $\omega_1, \omega_2 \neq 0$. The additional factor present
in Eq.~\eqref{unbinnedCovarianceGeneral} for the values
$\omega_1 - \omega_2 \in \Omega \mathbb{Z}$ and $\omega_1 + \omega_2 \in \Omega \mathbb{Z}$
has vanished due to the binning.

The variance of $S_{\text{bin}}(\omega)$ turns out to be
\begin{equation}
\operatorname{Var}\left[ S_{\text{bin}}(\omega) \right]
= V_{\text{bin}}(\alpha) \langle S_{\text{bin}}(\omega) \rangle^2
\label{varianceBinned}
\end{equation}
with the function
\begin{equation}
V_{\text{bin}}(\alpha) = 2 \frac{\Gamma(1+\alpha)^2}{\Gamma(1+2\alpha)} - 1.
\label{definitionV}
\end{equation}
For $0<\alpha<1$, the value of $V_{\text{bin}}(\alpha)$ is positive,
i.e., the variance of $S_{\text{bin}}(\omega)$ does not vanish and
the observable $S_{\text{bin}}(\omega)$ is a proper probability distribution. 
$V_{\text{bin}}(\alpha)$ describes the part of the fluctuation of the spectrum
which is connected with the weak ergodicity breaking, i.e., the
part that remains after binning. Correspondingly, for the ergodic
limit $\alpha \to 1$, the function $V_{\text{bin}}(\alpha)$ vanishes
($V(1) = 0$) and the observable $S_{\text{bin}}(\omega)$ does
not fluctuate.

In general, the whole spectrum is estimated. The question arises, how
the weak ergodicity breaking affect different frequencies. Is it possible
that one would estimate a wrong exponent $\beta$ of the $1/f^\beta$ noise?
This can be answered by using Eq.~\eqref{binnedCovarianceGeneral} to
calculate the correlation coefficient for all $0 < \alpha < 1$ and $\omega_1,\omega_2 \neq 0$:
\begin{equation}
\rho\left[ S_{\text{bin}}(\omega_1), S_{\text{bin}}(\omega_2) \right] = 1.
\label{binnedCorrelation}
\end{equation}
This implies that $S_{\text{bin}}(\omega_1)$ and $S_{\text{bin}}(\omega_2)$ are linearly coupled
\begin{equation}
S_{\text{bin}}(\omega_2) = \frac{\langle S_{\text{bin}}(\omega_2) \rangle}{\langle S_{\text{bin}}(\omega_1) \rangle} 
  S_{\text{bin}}(\omega_1).
\end{equation}
Therefore, as soon as one value of the spectrum is evaluated, the other values
follows from this one. In other words, the weak ergodicity breaking does only
affect a random prefactor which is the same for all frequencies.

\section{Numerical Verification}
\label{sectionNumericalSimulations}

In order to verify the presented analytical results we have
performed numerical realizations of the process described above.
To this purpose we have generated time series according to the
following procedure:
\begin{itemize}
\item Pick random numbers, $\tau_i$, from a totally asymmetric L{\'{e}}vy
$\alpha$-stable distribution with a given $\alpha$. These numbers
represent the waiting times. The random numbers were generated by
means of the procedure introduced in \cite{Weron01} with the scaling
parameter set to one.
\item Pick additional random numbers, $\chi_i$, from a normal
distribution. These numbers are the values of the process during
the respective time intervals.
\item The values $\chi_i$ are written into the entries of an
array, $A_j$, of total length $L$, for $j\in [\Delta t
\sum_{k=1}^{i-1}\tau_k, \Delta t \sum_{k=1}^{i}\tau_k[$. In all
simulations presented here we chose $\Delta t=0.25$, defining the
minimal resolution of the process.
\end{itemize}
The unbinned spectra, $S_\mathrm{ub}(\omega)$, can than be
determined straight forwardly by means of the fast Fourier
transform of the array $\mathbf{A}$. Of course, the spectra
obtained for these processes are discrete. In order to obtain the
binned spectra, $S_{\mathrm{bin}} (\omega,\Delta \omega)$, we
average $S_\mathrm{ub}$ over various consecutive frequencies and
assign the obtained value to the average of these frequencies.

\begin{figure}
\centerline{\includegraphics[width=75mm]{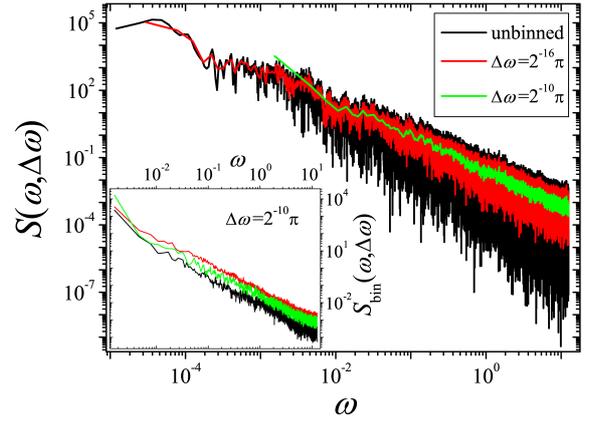}}
\caption{Power spectra corresponding to a single realization of
the waiting time process are plotted for $\alpha=0.5$ for the
unbinned case as well as for two different choices of the bin
size. In the inset we show spectra corresponding to different
realizations of the same process for a fixed choice of the binning
for $\alpha=0.5$. Note that the curves exhibit a shift with respect to
one another.\label{fig3}}
\end{figure}
In Fig.~\ref{fig3} we plot the unbinned spectrum of a typical
realization of a process of length $L=2^{21}$ for $\alpha=0.5$
together with the binned spectra for two choices of the binning,
$\Delta \omega=2^{-16} \pi$, and $\Delta \omega= 2^{-10}\pi$,
respectively. It should be observed how the spectra get smoother
when increasing the bin size, while the average slope of the
spectra is not changed by the binning. Deviations from the
power law behavior at small and large frequencies stem from the
finite size of array $\mathbf{A}$ and the minimal resolution of
the process, respectively.

The inset of Fig.~\ref{fig3} shows the binned spectrum with
$\Delta \omega= 2^{-10}\pi$ for various realizations of the
processes with length $L=2^{21}$ for $\alpha=0.5$. Note that,
while the exponent characterizing the decay of the spectrum does
not depend on the realization, the normalization factor of the
spectrum does. We would like to emphasize that the normalization
factor will neither converge by increasing the length of the
process nor by choosing larger bins. This probabilistic property
of the normalization factor is a consequence of the weak
ergodicity breaking.

\begin{figure}
\centerline{\includegraphics[width=75mm]{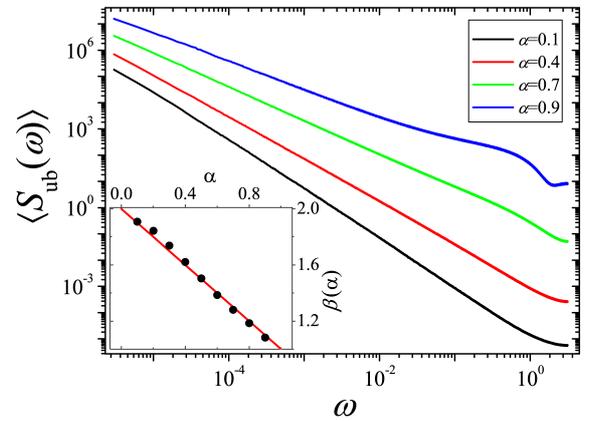}} \caption{The
averaged unbinned spectra corresponding to various choices of
$\alpha$ are plotted. Note that the spectra show an algebraic
decay over various decades of $\omega$ with an exponent $\beta$
depending on $\alpha$. The curves have been shifted for better
visibility. In the inset we compare the theoretical prediction for
the exponent $\beta$ (solid line) to the numerical findings
(dots).\label{fig4}}
\end{figure}
To verify our predictions with respect to the dependence of the
spectrum on $\alpha$, we have considered the unbinned spectrum
averaged over 10000 realizations of the process, $\langle
S_\mathrm{ub}(\omega)\rangle$. In Fig.~\ref{fig4} we plot $\langle
S_\mathrm{ub}(\omega)\rangle$ for processes of length $L=2^{21}$
and $\alpha=0.1,0.4,0.7,0.9$ (bottom to top), respectively. The
curves have been shifted with respect to one another for better
visibility. Note that the spectra show a $\sim 1/\omega^\beta$
behavior over various decades with an an exponent $\beta$ that
decreases with increasing $\alpha$. In the inset of
Fig.~\ref{fig4} we compare our prediction for the value of the
exponent $\beta(\alpha)=2-\alpha$ (solid line), see
Eq.~\eqref{expectationSpectrumOrigin}, to the values obtained
numerically by means of measuring the slope of the spectra in
log-log scale at the intermediate region. A fairly good agreement
is observed.

\begin{figure}
\centerline{\includegraphics[width=75mm]{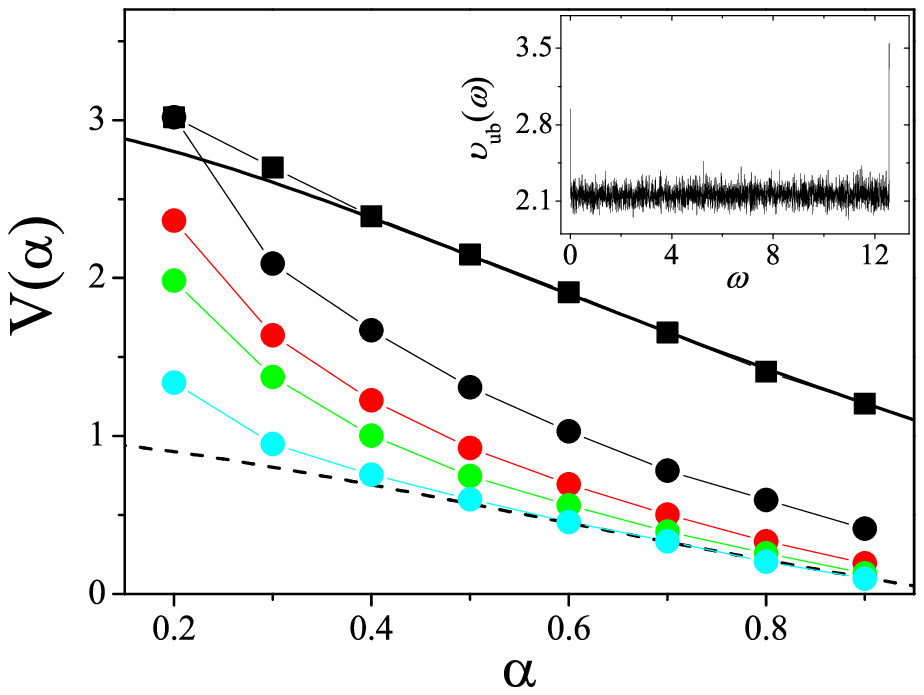}} \caption{The
theoretical predictions for $V(\alpha)$ are plotted for the
unbinned (solid line) and binned (dashed line) case. Numerical
results for the unbinned and the binned case are represented by
the square and circular dots, respectively. In the binned case,
various choices of process length have been considered,
$L=2^{13},2^{15},2^{17},2^{23}$ (top to bottom). The
inset shows the proportionality factor between the spectrum and the
square root of the variance for the unbinned case,
$\upsilon_{\mathrm{ub}}(\omega)=\mathrm{Var}[S_{\mathrm{ub}}(\omega)]/\langle
S_{\mathrm{ub}}(\omega)\rangle^2$, for $\alpha=0.5$. Note that it
is constant over various decades.\label{fig5}}
\end{figure}
Let us now check the predictions with respect to the variance of
the spectra. To calculate the variance of the unbinned spectra we
have considered processes of length $L=2^{21}$ and averaging was
carried out over 10000 realizations of the process. In accordance
with Eq.~\eqref{resultUnbinnedVariance} it is found that the
variance is proportional to the squared expectation value of the
spectrum over various decades with deviations at very high and
very small frequencies only (see inset of Fig.~\ref{fig5}). To
determine the proportionality factor we have averaged this factor
over the intermediate frequency region. In Fig.~\ref{fig5} the
theoretical prediction Eq.~\eqref{definitionVUnbinned} for the
proportionality factor $V_\mathrm{ub}(\alpha)$ (solid line) is
compared to the numerical results (square dots), for various
choices of $\alpha$. We find a reasonable correspondence with
deviations at very small $\alpha$. However, these deviations become
smaller when considering longer process lengths.

To verify the predictions for the asymptotic behavior of binned
spectra we have considered processes of various lengths,
$L=2^{13},2^{15},2^{17},2^{23}$, while the size of
the bins in frequency space was kept fixed at $\Delta \omega=
2^{-8}\pi$. Note that, by augmenting the process length while
keeping the bin size fixed, the number of frequencies of the discrete
spectrum falling into a specific bin increases. Again, the
averaging was carried out over 10000 realizations of each process.
In Fig.~\ref{fig5} the theoretical prediction for $V_{\text{bin}}(\alpha)$ for
the binned case (dashed line), see Eq.~\eqref{definitionV}, is
compared to the numerical data obtained for the binned processes
(circular dots). It should be observed that, as the length of the
process is increased (top to bottom), the values of $V_{\text{bin}}$ depart
from the predictions of the unbinned case (solid line) and
approach the predictions for the binned case (dashed line).

\begin{figure}
\centerline{\includegraphics[width=75mm]{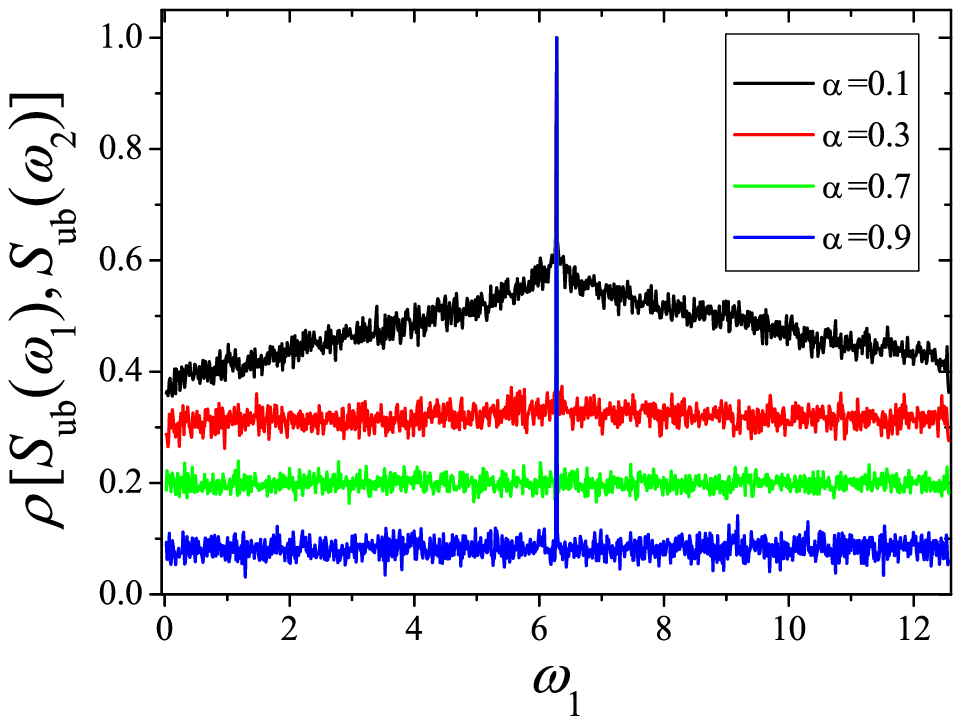}}
\caption{$\rho[S_\mathrm{ub}(\omega_1),S_\mathrm{ub}(\omega_2)]$
is plotted for $\omega_1=2\pi$ and various choices of $\alpha$.
The curves shift upwards as $\alpha$ is decreased. The peak
appears at $\omega_1=\omega_2$. For very small $\alpha$ the peak
extends over the whole frequency domain.\label{fig6}}
\end{figure}
Next, we have verified our predictions with respect to the
correlation coefficient of the spectra. For the unbinned case we
have again considered processes of length $L=2^{21}$ and averaged
over 10000 realizations. In Fig.~\ref{fig6}, the correlation
coefficient of the unbinned spectrum,
$\rho[S_\mathrm{ub}(\omega_1),S_\mathrm{ub}(\omega_2)]$, is
plotted for the fixed frequency $\omega_1=2\pi$ and various
choices of $\alpha$. Note that the correlation coefficient for not
too small $\alpha$ is indeed flat (apart from the trivial peak at
$\omega_1=\omega_2$) and that its value increases for decreasing
$\alpha$. For very small values of $\alpha$ the peak around
$\omega_1=\omega_2$ gets extended over the whole frequency domain
but this effect can be remedied by considering longer process
lengths.

\begin{figure}
\centerline{\includegraphics[width=75mm]{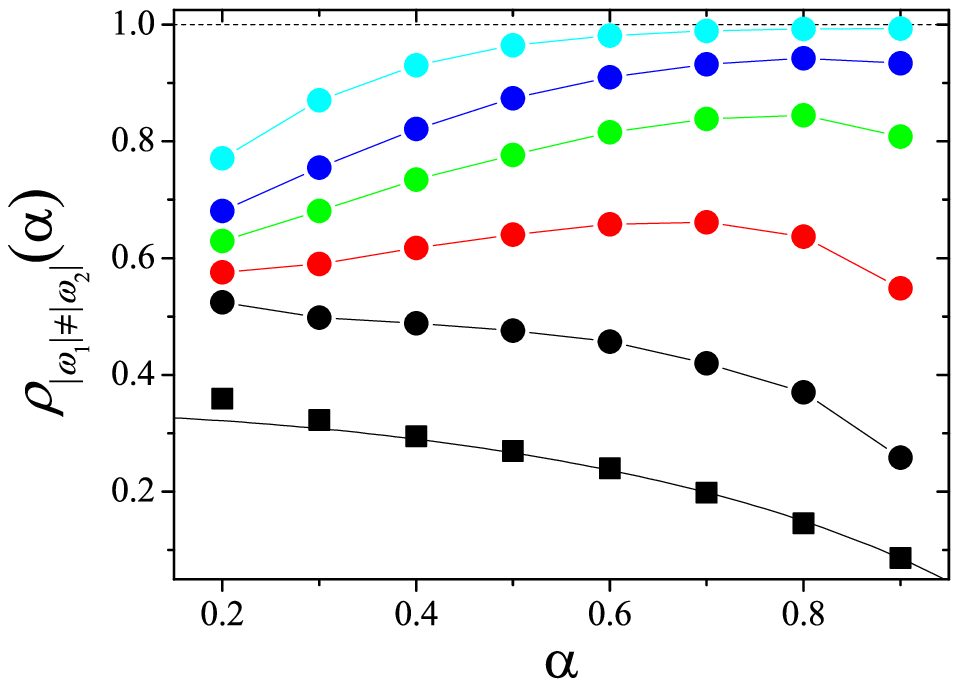}} \caption{The
theoretical predictions for $\rho_{|\omega_1|\neq
|\omega_2|}(\alpha)$ are plotted for the unbinned (solid line) and
binned (dashed line) case. Numerical results for the unbinned and
the binned case are represented by the square and circular dots,
respectively. In the binned case, various choices of process
length have been considered,
$L=2^{13},2^{15},2^{17},2^{19},2^{23}$ (bottom to top).
\label{fig7}}
\end{figure}
In Fig.~\ref{fig7} we compare the predicted value for the
correlation coefficient in the unbinned case (solid line),
$\rho_{|\omega_1|\neq |\omega_2|}(\alpha)=R(\alpha)$, see
Eq.~\eqref{resultUnbinnedCorrelation}, with the numerically
obtained data (squared dots). For the numerical data the
correlation coefficient was averaged over $2^{16}$ choices of the
pairs $\omega_1$ and $\omega_2$. Reasonable agreement between
theory and numerical data is found with deviations at small
$\alpha$ values stemming from the broadening of the peak around
$|\omega_1|=|\omega_2|$.

Finally, we repeated the analysis of the correlation coefficient
for the binned spectra. Again, we considered processes of various
lengths, $L=2^{13},2^{15},2^{17},2^{19},2^{23}$, while keeping
the bin size fixed in frequency space with $\Delta \omega=
2^{-8}\pi$. In Fig.~\ref{fig7} the theoretical prediction for the
correlation coefficient, $\rho_{|\omega_1|\neq
|\omega_2|}(\alpha)=1$, see Eq.~\eqref{binnedCorrelation}, is
plotted as dotted line while the numerical data is represented by
the circular dots. As the length of the process is increased, the
numerical values for $\rho_{|\omega_1|\neq |\omega_2|}$ depart
from the theoretical prediction for the unbinned case (solid line)
and approach the predicted line for the binned case (dashed line).

\section{Summary}

In this paper, we have determined the spectral properties of
a model by Rebenshtok and Barkai which shows weak ergodicity
breaking. The analytical results were verified by numerical
simulations. Near the origin, the spectrum shows a typical
$1/f^{2-\alpha}$ behavior with $\alpha$ characterizes the 
tail behavior of the waiting time. Using a single time series will result
in fluctuations of the unbinned spectrum which are also seen in the
ergodic case. Therefore, one commonly uses binning to determine
reliable values. While the fluctuations of the binned spectral
observable vanish in the ergodic case, this is not the case
for weak ergodicity breaking. However, the fluctuation does
only affect a common prefactor for the whole spectrum. 
Consequently, the measurement of the exponent $\beta$ of the $1/f^\beta$ behavior
is not hindered by the weak ergodicity breaking.

Recently, the emergence of universal fluctuation for processes
with weak ergodicity breaking has been discussed by He et al.~\cite{HeBurovMetzlerEtAl08},
Sokolov et al.~\cite{SokolovHeinsaluHanggiEtAl09} and Esposito et 
al.~\cite{EspositoLindenbergSokolov10}. It has been shown that for 
several time averaged observables as the average mean square displacement 
the fluctuations are described by a Mittag--Leffler distribution.
As can be infered from Eqs.~\eqref{varianceBinned} and \eqref{definitionV},
the first two moments of the binned spectrum are also in agreement with 
the assumption of a Mittag--Leffler distribution. However, it remains to
be seen in a future work if the binned spectrum does also belong
to this universality class.

The analytical results for the binned spectrum are valid in the limit
of infinite time series length. In a numerical realization of
this, binning means averaging only over a finite number of discrete frequencies
due to the finite process time.
It is therefore a nontrivial observation that these binned spectra
converge for numerical manageable process lengths
towards the theoretical prediction.

\section*{Acknowledgments}

We want to thank Eli Barkai und the referees for many helpful comments,
especially for pointing out the
connection to the universal fluctuations.

\appendix

\section{Calculation of the quadruple Laplace transform}
\label{appendixLaplace}

In this appendix, we derive the Laplace transform
Eq.~\eqref{LaplaceTransformFourTimes}. We want to mention that
it is possible to obtain and motivate this result by the more general
methods discussed in the derivation of Eq.~(71) in
\cite{NiemannKantz08}. First consider the function
\begin{equation}
g(\mathbf{t})
= \delta(t_1-t_2) (t_2-t_3)^{\alpha-1} \theta(t_2-t_3) \delta(t_3-t_4) t_4^{\alpha-1}
\end{equation}
with the Dirac delta $\delta(t)$ and the Heaviside step function
$\theta(t)$. Its quadruple Laplace transform can be directly
calculated:
\begin{equation}
\Laplacetransform[g(\mathbf{t})] = \frac{\Gamma(1+\alpha)^2}{\alpha^2}
\frac{1}{\Lambda_{\{1,2,3,4\}}^\alpha} \frac{1}{\Lambda_{\{1,2\}}^\alpha}.
\end{equation}
Integration yields
\begin{equation}
\begin{split}
&\int_0^{T_1} \dop t_1 \, \int_0^{T_2} \dop t_2 \, \int_0^{T_3} \dop t_3 \, \int_0^{T_4} \dop t_4 \,
  g(\mathbf{t}) \\
&\quad = \frac{1}{\alpha} \int_0^{\min(T_1,T_2,T_3,T_4)} \dop t \,
  \left(\min(T_1,T_2) - t\right)^\alpha t^{\alpha-1} \\
&\quad = \frac{1}{\alpha^2} \min(T_1,T_2)^\alpha \min(T_1,T_2,T_3,T_4)^\alpha \\
&\qquad \times \hypergeom \left(\alpha,-\alpha;1+\alpha; \frac{\min(T_1,T_2,T_3,T_4)}{\min(T_1,T_2)} \right).
\end{split}
\end{equation}
Eq.~\eqref{LaplaceTransformFourTimes} follows from the properties of the Laplace
transform with respect to integration.

\section*{References}

\bibliographystyle{elsarticle-num}

\begin{thebibliography}{10}
\expandafter\ifx\csname url\endcsname\relax
  \def\url#1{\texttt{#1}}\fi
\expandafter\ifx\csname urlprefix\endcsname\relax\def\urlprefix{URL }\fi
\expandafter\ifx\csname href\endcsname\relax
  \def\href#1#2{#2} \def\path#1{#1}\fi

\bibitem{PengHavlinStanleyEtAl95}
C.-K. Peng, S.~Havlin, H.~E. Stanley, A.~L. Goldberger, Quantification of
  scaling exponents and crossover phenomena in nonstationary heartbeat time
  series, Chaos: An Interdisciplinary Journal of Nonlinear Science 5~(1) (1995)
  82--87.

\bibitem{Weissman88}
M.~B. Weissman, $1/f$ noise and other slow, nonexponential kinetics in
  condensed matter, Rev. Mod. Phys. 60~(2) (1988) 537--571.

\bibitem{HoogeKleinpenningVandamme81}
F.~N. Hooge, T.~G.~M. Kleinpenning, L.~K.~J. Vandamme, Experimental studies on
  1/f noise, Reports on Progress in Physics 44~(5) (1981) 479--532.

\bibitem{BalascoLapennaTelesca02}
M.~Balasco, V.~Lapenna, L.~Telesca, $1/f^\alpha$ fluctuations in geoelectrical
  signals observed in a seismic area of southern italy, Tectonophysics 347~(4)
  (2002) 253 -- 268.

\bibitem{Bouchaud92}
J.~P. Bouchaud, Weak ergodicity breaking and aging in disordered systems,
  Journal de Physique I 2~(9) (1992) 1705--1713.

\bibitem{BelBarkai05}
G.~Bel, E.~Barkai, Weak ergodicity breaking in the continuous-time random walk,
  Physical Review Letters 94~(24) (2005) 240602.

\bibitem{BelBarkai06a}
G.~Bel, E.~Barkai, Weak ergodicity breaking with deterministic dynamics,
  Europhysics Letters 74~(1) (2006) 15--21.

\bibitem{MargolinBarkai06}
G.~Margolin, E.~Barkai, Nonergodicity of a time series obeying {L}\'{e}vy
  statistics, Journal of Statistical Physics 122~(1) (2006) 137--167.

\bibitem{PressTeukolskyVetterlingEtAl86}
W.~H. Press, S.~A. Teukolsky, W.~T. Vetterling, B.~P. Flannery, Numerical
  recipes in FORTRAN : the art of scientific computing, 2nd Edition, Cambridge
  University Press, Cambridge, 1986.

\bibitem{RebenshtokBarkai07}
A.~Rebenshtok, E.~Barkai, Distribution of time-averaged observables for weak
  ergodicity breaking, Physical Review Letters 99~(21) (2007) 210601.

\bibitem{RebenshtokBarkai08}
A.~Rebenshtok, E.~Barkai, Weakly non-ergodic statistical physics, Journal of
  Statistical Physics 133~(3) (2008) 565--586.

\bibitem{KammeyerKroschel89}
K.-D. Kammeyer, K.~Kroschel, Digitale Signalverarbeitungs - Filterung und
  Spektralanalyse mit MATLAB-\"{U}bungen, 6th Edition, Teubner, Wiesbaden,
  1989.

\bibitem{DrozhzhinovZavjalov80}
J.~N. Drozhzhinov, B.~I. Zav'jalov, Tauberian theorems for generalized
  functions with supports in cones, Mathematics of the USSR - Sbornik 36~(1)
  (1980) 75--86.

\bibitem{Drozhzhinov83}
Y.~N. Drozhzhinov, A multidimensional {T}auberian theorem for holomorphic
  functions of bounded argument and the quasi-asymptotics of passive systems,
  Mathematics of the USSR - Sbornik 45~(1) (1983) 45--61.

\bibitem{NiemannKantz08}
M.~Niemann, H.~Kantz, Joint probability distributions and multipoint
  correlations of the continuous-time random walk, Physical Review E 78~(5)
  (2008) 051104.

\bibitem{Feller71}
W.~Feller, An Introduction to Probability Theory and Its Applications - Volume
  II, 2nd Edition, John Wiley \& Sons, New York, 1971.

\bibitem{Weron01}
R.~Weron, Levy-stable distributions revisited: tail-index $>2$ does not exclude
  the levy-stable regime, International Journal of Modern Physics C 12~(2)
  (2001) 209--223.

\bibitem{HeBurovMetzlerEtAl08}
Y.~He, S.~Burov, R.~Metzler, E.~Barkai, Random time-scale invariant diffusion
  and transport coefficients, Physical Review Letters 101~(5) (2008) 058101.

\bibitem{SokolovHeinsaluHanggiEtAl09}
I.~M. Sokolov, E.~Heinsalu, P.~H\"{a}nggi, I.~Goychuk, Universal fluctuations
  in subdiffusive transport, Europhysics Letters 86~(3) (2009) 30009.

\bibitem{EspositoLindenbergSokolov10}
M.~Esposito, K.~Lindenberg, I.~M. Sokolov, On the relation between event-based
  and time-based current statistics, Europhysics Letters 89~(1) (2010) 10008.

\end{thebibliography}

\end{document}